\documentclass{article}
\usepackage{ASRU_Template/spconf,amsmath,graphicx}
\usepackage[dvipsnames]{xcolor}
\usepackage{url}
\usepackage{balance}

\title{AC-VC: Non-parallel Low Latency Phonetic Posteriorgrams Based\\Voice Conversion}
\name{Damien Ronssin$^{1,2}$, Milos Cernak$^1$}

\address{
  $^1$Logitech Europe S.A., 1015, Lausanne, Switzerland\\
  $^2$École Polytechnique Fédérale de Lausanne (EPFL), 1015, Lausanne, Switzerland}
  
\begin{document}

\maketitle
\begin{abstract}
This paper presents AC-VC (Almost Causal Voice Conversion), a phonetic posteriorgrams based voice conversion system that can perform any-to-many voice conversion while having only 57.5 ms future look-ahead. The complete system is composed of three neural networks trained separately with non-parallel data. While most of the current voice conversion systems focus primarily on quality irrespective of algorithmic latency, this work elaborates on designing a method using a minimal amount of future context thus allowing a future real-time implementation. According to a subjective listening test organized in this work, the proposed AC-VC system achieves parity with the non-causal ASR-TTS baseline of the Voice Conversion Challenge 2020 in naturalness with a MOS of 3.5. In contrast, the results indicate that missing future context impacts speaker similarity. Obtained similarity percentage of 65\% is lower than the similarity of current best voice conversion systems.
\end{abstract}

\section{Introduction}
Voice conversion (VC) consists in transforming a speech audio signal so that the voice of the original speaker is changed into the one of a target speaker, while preserving the linguistic content. VC opens new avenues of creativity and productivity; for example, Blue Yeti X WoW Edition\footnote{\url{www.bluemic.com/en-us/products/yetixwow/}} offers real-time voice effects for someone to sound like any World of Warcraft character. Would it be possible to propose a similar effect performing real-time voice conversion ?

Nowadays, VC methods are based on machine learning. Older deep learning approaches required parallel data~\cite{todavcc2016}, i.e., a training dataset with parallel utterances from source and target speakers, to then train a sequence-to-sequence model mapping source acoustic features into target ones. As obtaining large amounts of parallel data is challenging, more recent works have concentrated on building VC systems trainable with non-parallel data. Such systems rely on an intermediate speaker-independent speech representation like text \cite{baselinepaper}, phonetic posteriorgrams (PPGs)\cite{sun2016phonetic,Zheng2020Casia} or disentangled latent space from an auto-encoder \cite{qian2019autovc, barbany2020fastvc}. This intermediate representation should carry only the linguistic content from the input utterance and nothing about the source speaker's voice characteristics. It will then be used to generate the voice-converted audio sample with a system trained with some target speaker data. 

In the voice conversion literature, most published papers focus on designing VC systems yielding the highest output quality but rarely on creating real-time VC techniques. Only few recent works such as \cite{Arakawa2019realtime} and \cite{saeki2020realtime} have been pursued to build real-time VC methods. These approaches are briefly described in the next section. Nonetheless, having such real-time systems would dramatically widen the range of applications of voice conversion. 

In our previous work \cite{barbany2020fastvc}, we have designed the FastVC voice conversion algorithm that runs four times faster than real-time, i.e., that synthesizes 1s of converted speech in 0.25s on a CPU. We used this algorithm in our participation (team T15) in the Voice Conversion Challenge (VCC) 2020 \cite{vcc2020}. The VCC is a bi-annual challenge where scientists can submit their voice conversion system to be evaluated and compared with others on the same dataset. The 2020's session consisted of two VC tasks, an intra-lingual one with semi-parallel data and a cross-lingual one. FastVC participated only in the cross-lingual VC task and outperformed two of the three Challenge baselines in terms of naturalness. 

However, FastVC can't be used in real-time as it is not causal. In this work, we address this issue by proposing AC-VC (Almost Causal Voice Conversion), a voice conversion system operating with a reasonable algorithmic latency, i.e., using a look-ahead of maximum 60 ms. As strictly causal systems only use past and current input samples to generate the current output sample, the proposed system is referred to as ``almost causal" to specify that some additional future context is also used. There is no published system trainable on non-parallel data with such a small latency to the best of our knowledge. 

AC-VC is composed of 3 separately trained neural networks. The first is an LSTM-based acoustic model that outputs Phonetic PosteriorGrams (PPGs). The second one is the conversion model that takes as input PPGs concatenated with one-hot target speaker embedding and fundamental frequency (F0) values and outputs converted acoustic features. The last one is the LPCNet, a speech vocoder designed by Mozilla teams \cite{valin2019lpcnet}, that takes as input the converted acoustic features and synthesizes the voice-converted speech. The proposed AC-VC system has a similar architecture to the CASIA Voice conversion system~\cite{Zheng2020Casia} that ranked second overall at the VCC 2020 (team T29)~\cite{vcc2020}, but it is low-latency. LPCNet and the CASIA voice conversion systems are presented in the next section.

\section{Related work}

\subsection{Real-time VC}
In \cite{Arakawa2019realtime}, a 50 ms latency voice conversion method is presented. It uses a deep neural network (DNN) to map source mel-frequency cepstral coefficients (MFCCs) and source F0 into target MFCCs, band-averaged aperiodicity and target F0. Then, recursive maximum likelihood parameter generation (R-MLPG) is applied to the previously converted MFCCs and finally, converted speech is synthesized with the WORLD vocoder \cite{morise2016world}. 

A different real-time VC approach based on spectral differentials combined with a DNN is presented in \cite{saeki2020realtime}. This system performs voice conversion without vocoding, i.e., no waveform is synthesized, the source audio is directly modified with a filter, reducing drastically the computational load. Consequently, this system runs faster than real-time on a CPU with only a few tens of ms latency.

While the use of parallel data is not directly indicated in the above-mentioned papers \cite{Arakawa2019realtime, saeki2020realtime}, one can deduce regarding the training phase descriptions that parallel utterances seems indeed required. The need for such data is a strong constraint that limits the number of possible source/target speaker pairs.

In this work, we hypothesized that PPGs-based VC could perform low-latency, similarly, as streaming speech recognition systems are almost real-time thanks to causal acoustic models (PPGs extractors). Besides, we designed the AC-VC system intending to train it from non-parallel data.

\subsection{LPCNet}
\label{lpcnet_section}

Causal PPGs extraction is mandatory but not sufficient. The major delay's cause is often speech vocoding. In this work, we chose to use LPCNet: it is a neural speech synthesizer adapted from WaveRNN \cite{kalchbrenner2018wavernn} that leverages linear prediction to perform speech synthesis. Instead of generating the final audio waveform directly, LPCNet synthesizes the excitation signal that is spectrally flat and thus simpler to model. Finally, the excitation is added to the linear prediction to obtain the audio waveform. 

LPCNet takes as input 18 Bark-scale cepstral coefficients (BSCCs), the pitch, and the pitch correlation for every 10 ms time frame.
The complete model comprises two neural networks: the frame rate network and the sample rate network. The former network, composed of two convolutions followed by two fully connected layers, takes the BSCCs as input and outputs a 128-dimensional conditioning vector. As indicated in its name, this network runs for each frame, i.e., every 10 ms. The sample rate network executes for each sample, so 16000 times per second to generate 16 kHz audio. It takes as input the conditioning vector from the frame rate network and the linear predicted sample, the previously generated excitation sample, and the previous output sample. It is composed of two Gated Recurrent Units (GRU) followed by a dual fully connected layer with a softmax activation. The output is a probability distribution over 256 values (8 bits) from which the following excitation sample is randomly sampled. 

Note that the sample rate network is causal and that the frame rate network has only 30 ms future look-ahead because of the two convolutions with a total receptive field of 5 frames. So the analysis re-synthesis with LPCNet can be done with only 30 ms latency. 

Also, LPCNet is an auto-regressive model as the previous output sample is given back as input to the network to generate the next one. It is known that large auto-regressive neural speech synthesizers such as WaveNet \cite{oord2016wavenet} can't generate speech in real-time as it is impossible to parallelize the generation of multiple audio samples. However, LPCNet~\cite{valin2019lpcnet} can run faster than real-time on a single CPU core. This is achieved thanks to the small model size and important optimizations such as the use of block-sparse matrices in a GRU layer along with efficient vectorizations. 

\subsection{CASIA VC}
\label{sec:casia}
The CASIA voice conversion system \cite{Zheng2020Casia} was submitted to the VCC 2020 and ranked in the very best teams on the two tasks, both in terms of naturalness and speaker similarity. It uses PPGs as speaker-independent speech representation. Thus, it is composed of an acoustic model to obtain the PPGs from source speech, a conversion model to get converted acoustic features from PPGs, and a vocoder to synthesize the output speech. 

More precisely, the acoustic model is a TDNN-LSTM network built with the Kaldi toolkit~\cite{povey2011kaldi}. It takes as input 40-D filter-bank features computed on 25 ms frames taken every 10 ms. A 512-D representation is then obtained for each frame by taking the output of the last LSTM layer. This 512-D representation is referred to as phonetic posteriorgrams (PPGs) in the CASIA VC paper \cite{Zheng2020Casia}, while it is not a probability vector and does not sum up to one. For simplicity and consistency, this designation is kept in this paper.

The conversion model is based on the CBHG (1-D Convolution Bank + Highway network + bidirectional GRU) module from Tacotron~\cite{wang2017tacotron}. The input comprises: (i) an input sequence of previously computed 512-D PPGs concatenated with (ii) one hot target speaker embeddings, (iii) target F0 values, (iv) voiced/unvoiced flag (VUF), and (v) band aperiodicity. The three latter values are obtained using tools from the WORLD vocoder \cite{morise2016world}. Finally, the conversion model generates a sequence of 30-D bark scale cepstral coefficients along with F0 and F0 autocorrelation values. 

Lastly, the vocoder is an adapted version of LPCNet. It takes as input the sequence as mentioned above of 32-D acoustic features and synthesizes 24 kHz audio waveform. 

The authors fine-tuned both the conversion model and LPCNet for each target speaker to improve their voice conversion quality. The authors reported that it allowed obtaining results that better match the target speaker's voice and prosody. The fine tuning process is described in more details in section \ref{fine_tuning_section}.

\section{Proposed method}

The CASIA team's submission in the VCC 2020 uses the LPCNet as a vocoder and satisfies our low latency requirement for speech synthesis. However, the system's other modules, namely the acoustic and conversion models, are not low latency, as they both use a large future context.  Different low latency acoustic models, including TDNN-LSTM models, are presented and compared in \cite{peddinti2017lowlatencyacousticmodel}. The lowest achieved latency is 70 ms, which is already too high for real-time voice conversion; one needs to add further the latency introduced both by the conversion model and the vocoder. The CBHG module uses bi-directional recurrent layers and is thus not suited for streaming or low latency tasks. 

\begin{figure}[h]
  \centering
  \includegraphics[trim=0 0 0 0, clip,width=\linewidth]{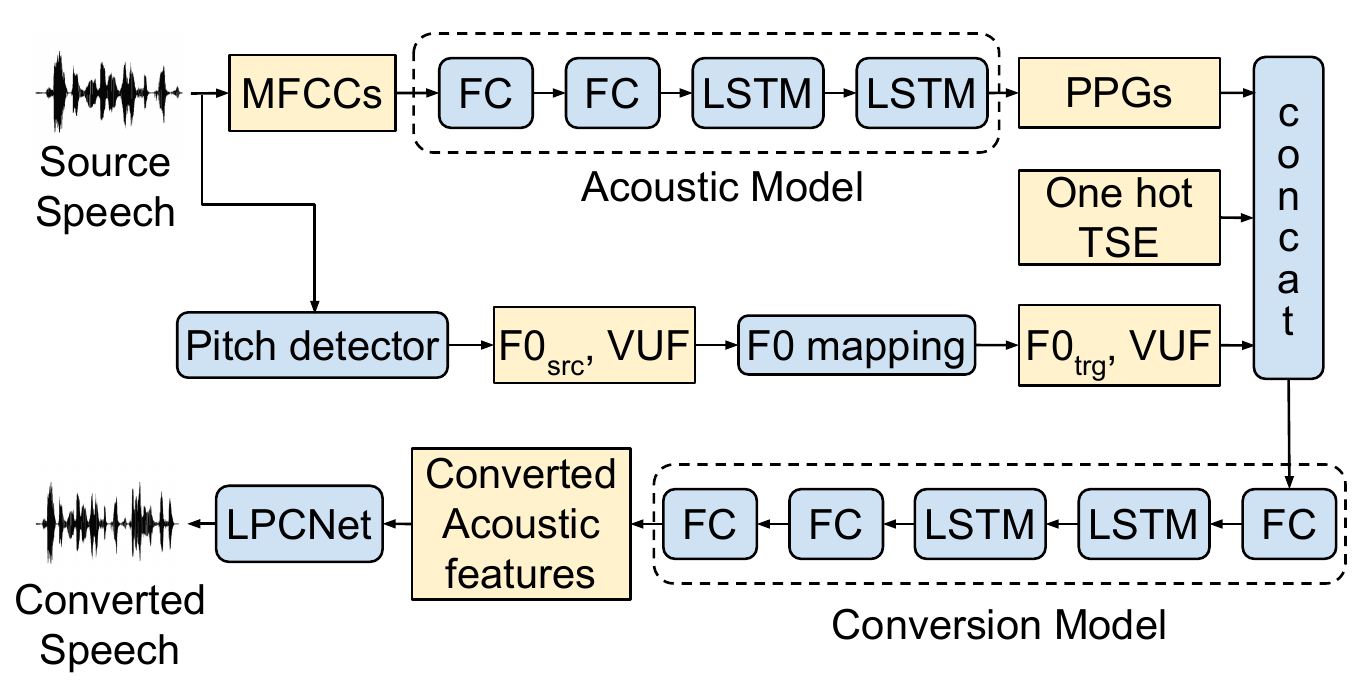}
  \caption{Schematic diagram of AC-VC. Yellow blocks represent signals while blue blocks represent operators. FC stands for Fully Connected Layer, TSE for Target Speaker Embedding, F0 for fundamental frequency and VUF for voiced/unvoiced flag.}
  \label{fig:VC_system}
\end{figure}

In our proposal, we decided to keep the overall architecture similar to the CASIA's VC system. Still, we incorporate new simple acoustic and conversion models that allow having a total latency of less than 60 ms. Figure \ref{fig:VC_system} shows a schematic overview of the proposed VC system.

\subsection{Acoustic model}

The acoustic model's role is to extract from the source audio a sequence of speaker-independent speech representation, in our case, phonetic posteriorgrams.

\subsubsection{Input features}

We chunk the source into overlapping frames of 25 ms with a hop size of 10 ms.  Then for each frame, 13-D Mel Frequency Cepstral Coefficients (MFCCs) are extracted along with their $\Delta$ and $\Delta\Delta$ values that give a sequence of 39-D acoustic features.

\subsubsection{Model structure and training}

The acoustic model is obtained by removing the last fully connected layer of a phoneme classifier. This classifier tries to predict a spoken phoneme in the current frame. It has the following simple structure: 2 time-distributed fully connected layers with ReLU activation followed by two uni-directional LSTMs and a final time-distributed fully connected layer with softmax activation. We train the model with the cross-entropy loss. In this configuration, the model is in ``streaming mode", i.e., the model tries to predict the phoneme uttered in the current frame while having no access to any future context. To relax this constraint and thus obtain a better classification accuracy, the target phoneme sequence has been shifted by one during training. The model learns to predict the phoneme of the previous frame. However, this manipulation introduces one frame latency in the prediction.

The acoustic model does not contain the last fully connected layer of the phoneme classifier as shown in Figure \ref{fig:VC_system}. Thus, the PPGs refer to the last LSTM layer's output and are 512-D similarly as~\cite{Zheng2020Casia}.

\subsection{Conversion model}

The conversion model's role is to generate a sequence of converted acoustic features for LPCNet to synthesize target speaker speech with the same linguistic content as the source speech. As we have seen in Section~\ref{lpcnet_section}, LPCNet requires 18 BSCCs, current F0, and F0 autocorrelation for each 10 ms frame and will generate speech accordingly. 

\subsubsection{Model inputs}

The conversion model takes input 512-D PPGs concatenated at each time frame with the target F0, the voiced/unvoiced flag (VUF), and the  N-D one hot target speaker embedding (with N possible target speakers). The target F0 is obtained via a transformation of the source F0 that maps the source speaker's F0 range to that of the target speaker. This transformation is referred as $F0$ mapping in Figure \ref{fig:VC_system} and is described with more details in~\cite{Zheng2020Casia}. It requires the knowledge of the F0 mean and variance of both the source and target speakers.

\subsubsection{Model structure}

The conversion model has a similar structure as the acoustic model. Indeed, it is composed sequentially of one time-distributed fully connected layer with ReLU activation, two uni-directional LSTM layers, and two time-distributed fully connected layers with ReLU and no activation respectively. A simple L1 loss was applied during training. Similarly, as with the acoustic model, the target sequence has been time-shifted by one so that the model predicts the converted acoustic features of the previous frame. This simplifies the network's learning task as more future context is available but adds one more frame of latency.

\subsubsection{Training and fine tuning} \label{fine_tuning_section}

The conversion model's training does not require any parallel data. During training, the model tries to predict the acoustic features of the source utterance given phonetic posteriorgrams computed by the acoustic model, the source speaker one hot embedding, and F0 values. Thus, no voice conversion occurs at training time. Also, the acoustic and conversion models are not trained jointly as the acoustic model would risk becoming speaker-dependent. After this training phase, one obtains the average conversion model, i.e., the conversion model trained with multi-speaker data.

To further improve the VC system's performance, both the average conversion model and average LPCNet are fine-tuned for a specific target speaker. This operation consists in continuing the training of the average model using only speech data from that specific speaker. This is usually done with a smaller learning rate to avoid obtaining a model too different from its average version.

\subsection{Implementation details}

The models were implemented, trained, and tested using Tensorflow and Keras. For the acoustic model, the LibriSpeech clean dataset \cite{panayotov2015librispeech} was used with phoneme alignments from \cite{loren_lugosch_alignments}. The VCTK dataset \cite{veaux2016_vctk} composed of 109 speakers was used for the conversion model training. Both models were trained with a batch size of 32, the Adam optimizer with a learning rate of $0.001$, and have approximately 2 million parameters. 

For LPCNet, we used the open-source Python and C codes provided by the authors\footnote{\url{https://github.com/mozilla/LPCNet}} for fine-tuning and inference, respectively. Additionally, the built-in pitch detector of LPCNet was used to extract F0 and VUF from the input utterance. 

Given that both the acoustic and conversion models have one frame look-ahead, that LPCNet has two frames look-ahead, and that frames are 25 ms with a hop size of 10 ms, the system's overall latency is 57.5 ms. Our experiments showed that using both the acoustic and the conversion models in ``streaming mode`` (with no future frame look-ahead) is possible. That reduces the latency to 37.5 ms and provided a similar voice conversion quality. However, this configuration was not properly evaluated so it is not discussed more in this paper.

Note that the LPCNet pitch detector introduces approximately 10 ms of additional latency in the system as it uses F0 trajectory optimization. This additional latency is not accounted for in the latency mentioned above of 57.5 ms as another lower-latency pitch detector may be used instead.

\subsection{Baselines}
To assess the performance of the proposed AC-VC system, two different voice conversion systems serve as baselines. The first and main baseline is the ASR-TTS system also used as baseline in the VCC2020 \cite{vcc2020} (T22). It is composed of a transformer based ASR, of a multispeaker x-vector Transformer TTS model, and of parallel WaveGAN (PWG) vocoder \cite{yamamoto2020pwg}. It is described in more details in \cite{baselinepaper}. The open-source implementation of this system from ESPnet\footnote{\label{footnote_github_espnet}\url{https://github.com/espnet/espnet}}~\cite{watanabe2018espnet} was used generate voice converted audio samples evaluated in this work.

As AC-VC is a low latency adaptation of the CASIA VC system,  the latter constitutes logically our second baseline. Note that this system was not reproduced in this work as no open source implementation was available. Consequently, the results of the VCC2020 \cite{vcc2020} (T29) were used instead.

\section{Experimental setup}

We evaluated the proposed AC-VC system's performance with a subjective test that we organized on the Amazon Mechanical Turk (AMT) platform. This listening test contained natural speech samples and voice-converted speech samples from both AC-VC and the ASR-TTS baseline from the VCC2020.

We chose six target speakers from the VCTK test set (p252, p256, p268, p301, p307, and p316), and fine-tuned six models for both the proposed and the baseline systems. They have been chosen in order to obtain a balanced target speaker set in term of genders, pitch ranges and dialects. Similarly, 4 source speakers have been chosen (p267, p270, p299 and p311) making a total of 24 source/target speakers conversion pairs for the subjective evaluation.

The fine-tuning process of AC-VC is described in section \ref{fine_tuning_section}. Simultaneously, for the baseline, the instructions given in the ESPnet recipe ``vcc2020/vc1\_task1" were carefully followed, i.e., only the TTS module was fine-tuned with target speaker data, while the ASR and PWG modules were left as is.

The p.808 toolkit\footnote{\url{https://github.com/microsoft/P.808}} from Microsoft \cite{naderi2020p808microsoft} was used to design the evaluation on Amazon Mechanical Turk. More precisely, the p835 template was adapted for voice conversion, as follows. 

The subjective evaluation was designed similarly as in the Voice Conversion Challenge 2020~\cite{vcc2020}. The listeners first rated naturalness, rating two audio samples, a natural speech one and a voice-converted one, on a scale from 1 to 5 (\textit{Bad, Poor, Fair, Good, Excellent}). In the second test, the listeners rated speaker similarity, listening to the same audio samples, and rating the pair for speaker similarity on a scale from 1 to 4, meaning \textit{different speakers (sure), different speakers (not sure), same speaker (not sure), same speaker (sure)}. A total number of 26 unique listeners participated in the evaluation. More than 100 and 80 audio samples per VC system have been rated for Mean Opinion Score (MOS) and speaker similarity, respectively. Each audio sample or pair has been rated by at least four different listeners.

\section{Results and discussion}

\begin{figure}[h]
  \centering
  \includegraphics[trim=20 15 15 25, clip, width=\linewidth]{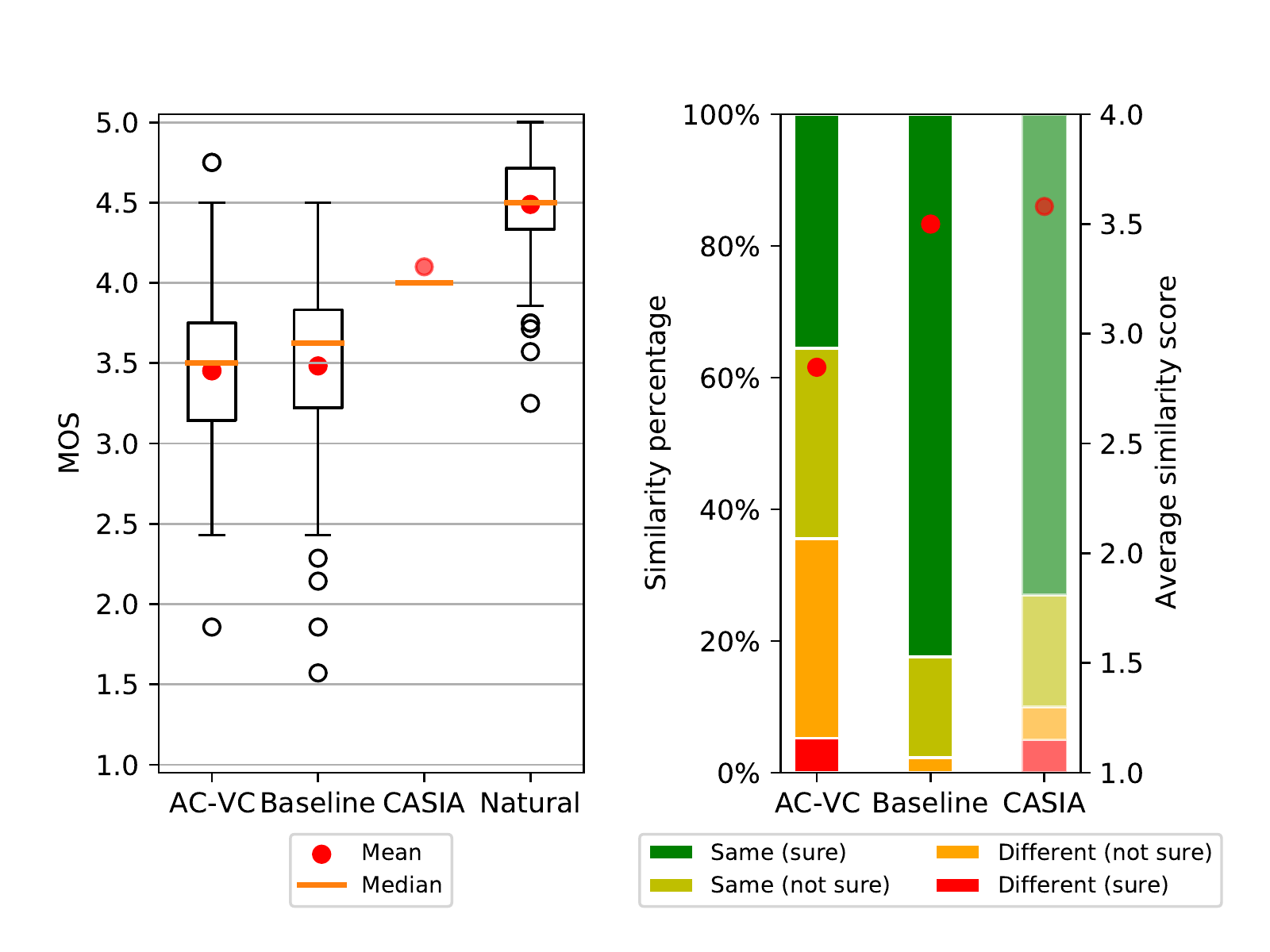}
  \caption{On the left: MOS ratings for the three VC systems (AC-VC, ASR-TTS Baseline and CASIA VC) and the natural speech samples. Mean values (red dots) have been added to the standard boxplot. On the right: Speaker similarity rating distributions (colored bars, left scale) and average similarity scores (red dots, right scale), for the three VC systems. The CASIA VC system was not evaluated in our subjective test, its results were taken from VCC 2020 \cite{vcc2020}.}
  \label{fig:results}
\end{figure}

The results of the subjective evaluation are presented in Figure \ref{fig:results}. As the obtained results for our reproduction of the ASR-TTS baseline of VCC2020 are very similar to those obtained in the VCC2020's subjective evaluation \cite{vcc2020}, our results can be compared accurately with those of the VCC2020. Consequently, the CASIA VC system results from VCC2020's evaluation are included in Figure \ref{fig:results} with attenuated colors to indicate the different origin of the data. Note that as the MOS scores were processed slightly differently in \cite{vcc2020}, only the mean and median are reported for the CASIA system.

  Some AC-VC and ASR-TTS voice conversion examples can be listened online\footnote{\url{https://damrsn.github.io/AC-VC/}}. 

On the left side of Figure \ref{fig:results}, one can see the naturalness rating distributions for the proposed system, the ASR-TTS baseline, CASIA VC and natural speech samples from VCTK. While CASIA VC and natural samples achieve naturalness MOS higher than 4, AC-VC and the ASR-TTS baseline perform equally well with both MOS being close to 3.5. A t-test between the AC-VC and baseline MOS reported $t = -0.41$ and $p = 0.68$. Consequently, the null hypothesis of equal averages can't be rejected, i.e., no significant difference between the AC-VC and the ASR-TTS baseline MOS can be established. 

This result represents the main achievement of this work; it demonstrates the possibility to obtain natural voice converted speech with limited future context. As said before, small algorithmic latency is a necessary condition to consider a possible real-time implementation. For the latter, computational considerations are of equal importance, but these are beyond this paper's scope and have been studied in a previous work~\cite{barbany2020fastvc}.

On the right side of Figure \ref{fig:results}, one can see the speaker similarity rating distribution for each VC system over the four values outlined above. The average similarity score is also indicated. Regarding those criteria, the ASR-TTS baseline and CASIA VC perform significantly better than the proposed system with similarity percentages of 98\% and 90\% against 65\%. The similarity percentage is defined as the proportion of the aggregate of \textit{same speaker (sure)} and \textit{same speaker (not sure)} ratings. 

We hypothesize that a part of this performance difference is due to the missing future context of AC-VC. In general, modifying source speech prosody is more challenging for a low latency system. For example, the speech rate can't be modified at all in such setting. Similarly, the word emphasis and the F0 trajectory are somehow fixed because future context is needed to perform a meaningful and consistent transformation. On the contrary, an offline system such as an ASR-TTS can freely reshape the source signal and perfectly match the target speaker's prosody. Still, AC-VC's \textit{Different (sure)} rating proportion is as low as that of CASIA VC at 5\%, indicating that the source voice almost always get transformed toward the target one.

The acoustic and conversion models' simplicity implies that further quality improvements, both in term of naturalness and speaker similarity, are possible with further exploration of the models' architecture. A more sophisticated acoustic model could indeed allow to obtain better and more speaker independent PPGs. Also, adding the band aperiodicity to the conversion model's input and using a greater number of acoustic features as input to LPCNet, similarly as in \cite{Zheng2020Casia}, could yield higher quality converted acoustic features and synthesized speech.

\section{Conclusions}

In this work, we have introduced AC-VC (Almost Causal Voice Conversion), a low latency phonetic posteriorgrams based voice conversion system trained with non-parallel data. AC-VC uses a future look-ahead of 57.5 ms to generate voice-converted audio samples, allowing a possible real-time implementation. It can perform any-to-many voice conversion as the only required information about the source speaker is their mean and variance pitch values, which can easily be computed on the fly. We have demonstrated that AC-VC is the novel VC system trainable with non-parallel data achieving such a small latency as other existing real-time voice conversion systems use parallel data for training.

According to a subjective evaluation organized in this work, AC-VC performs equally well as the ASR-TTS baseline of the Voice Conversion Challenge 2020 in terms of naturalness with a MOS of 3.5. However, the latter yields significantly higher performances in terms of speaker similarity, with a similarity percentage of 98\% against 65\% for the proposed system. Reducing this similarity gap thus paves our future work.

\section{Acknowledgements}
We would like to thank Natalia Nessler, Pablo Mainar and Tibor Vass for their precious help in the design of the subjective test.

\bibliographystyle{IEEEbib}
\bibliography{mybib}
\balance
\end{document}